# Enhancing COVID-19 Diagnosis through Vision Transformer-Based Analysis of Chest X-ray Images

Sultan ZAVRAK[1]

Sultan ZAVRAK

[1]Department of Computer Engineering, Duzce University, Duzce/TURKEY

sultanzavrak@duzce.edu.tr

***ORCID ID:*** 0000-0001-6950-8927

***Corresponding Author:**

Sultan ZAVRAK

Department of Computer Engineering
Faculty of Engineering, Duzce University
Konuralp Campus, 81620, Duzce / Turkey
Phone: +90(380)5421036/4730
E-mail: sultanzavrak@duzce.edu.tr , sultanzavrak@gmail.com

# Enhancing COVID-19 Diagnosis through Vision Transformer-Based Analysis of Chest X-ray Images

## Abstract

The advent of 2019 Coronavirus (COVID-19) has engendered a momentous global health crisis, necessitating the identification of the ailment in individuals through diverse diagnostic modalities. Radiological imaging, particularly the deployment of X-ray imaging, has been recognized as a pivotal instrument in the detection and characterization of COVID-19. Recent investigations have unveiled invaluable insights pertaining to the virus within X-ray images, instigating the exploration of methodologies aimed at augmenting diagnostic accuracy through the utilization of artificial intelligence (AI) techniques. The current research endeavor posits an innovative framework for the automated diagnosis of COVID-19, harnessing raw chest X-ray images, specifically by means of fine-tuning pre-trained Vision Transformer (ViT) models. The developed models were appraised in terms of their binary classification performance, discerning COVID-19 from Normal cases, as well as their ternary classification performance, discriminating COVID-19 from Pneumonia and Normal instances, and lastly, their quaternary classification performance, discriminating COVID-19 from Bacterial Pneumonia, Viral Pneumonia, and Normal conditions, employing distinct datasets. The proposed model evinced extraordinary precision, registering results of 99.92% and 99.84% for binary classification, 97.95% and 86.48% for ternary classification, and 86.81% for quaternary classification, respectively, on the respective datasets.

## 1. Introduction

In December 2019, an incipient epidemic of the coronavirus disease (COVID-19) was initially documented in Wuhan, China, precipitating a rapid dissemination within the nation and subsequently across global borders. Designated as a worldwide pandemic by the World Health Organization (WHO) [1], this affliction stems from severe acute respiratory syndrome coronavirus 2 (SARS-CoV-2). COVID-19 manifests a disconcerting manifestation, wherein research findings evince a mortality rate exceeding 60% upon progression to severe or critical stages of the malady [2], [3]. Predominant factors contributing to fatality encompass extensive alveolar degradation and progressive respiratory insufficiency [4]. Moreover, the transmission of SARS-CoV-2 occurs stealthily, even in the absence of overt symptoms, heightening the urgency for expeditious and precise COVID-19 screening and diagnosis, early intervention strategizing, interception of transmission pathways, and formulation of clinical modalities to ameliorate prognostic outcomes [5].

In recent years, the utilization of machine learning (ML) methodologies in the realm of medicine for the purpose of automated diagnosis has gained considerable traction as a complementary instrument for clinicians [6]–[8]. Deep learning (DL), a well-established subdomain within the field of artificial intelligence (AI), assists the production of end-to-end models that leverage input data to yield desired outcomes, thereby obviating the requirement for manual feature extraction [9], [10]. DL methods have been efficaciously employed for diverse problem domains, encompassing the detection of arrhythmias [11], classification of brain diseases [12], identification of breast cancer [13], and discernment of pneumonia from chest X-ray images [14], [15]. The rapid proliferation of the COVID-19 pandemic has underscored the urgency for expertise in this realm, intensifying interest in the creation of AI-driven automated detection systems. Given the scarcity of radiologists, the task of staffing every hospital with proficient clinicians poses a formidable challenge. Consequently, the deployment of straightforward, accurate, and expeditious AI models may prove beneficial in ameliorating this predicament and providing timely aid to patients. While radiologists hold a pivotal role in this domain owing to their extensive expertise, AI technologies in radiology can serve as valuable aids in facilitating precise diagnoses [16].

In order to aid in the identification of SARS-COV2 infection, Hassantabar et al. [17] employed a Convolutional Neural Network (CNN) enhanced with the neural network layer using the Softmax activation function. The DarkCovidNet architecture, developed by Ozturk et al. [18], incorporates 17 convolutional layers and employs diverse filtering techniques at each layer. This methodology enables accurate discrimination between binary and multi-class classifications. Pertaining to the diagnosis of COVID-19 via chest radiographs, Pandit et al. [19] undertook the refinement of the VGG-16 network model architecture. In the evaluation of X-ray images, Apostolopoulos and Mpesiana [20] conducted a comprehensive investigation encompassing MobileNet, Xception, Inception ResNetv2, Inception, and VGG-19. Ismael and Sengur et al. [21] employed Support Vector Machine (SVM) and Resnet50 characteristics derived from X-ray data to establish a model for COVID-19 identification. Employing aggregation via contrastive loss, Li et al. [1] studied a multi-task contrastive learning approach to perform automated detection of COVID-19 infections. Rashid et al.'s [22] design, known as AutoCovNet, employs a two-stage deep CNN architecture. This design incorporates an encoder-decoder-based auto-encoder network, an encoder-fusion network, a sophisticated feature-fusion technique, and an EfficientNet-B4 network to distinguish COVID-19 in chest X-ray images. Gayathri et al. [23] suggested the utilization of feedforward neural networks for classification purposes, employing a reduced Autoencoder (AE) for dimensionality reduction, and leveraging pre-trained CNNs to perform feature extraction for COVID-19 detection with computer help. Sundaram et al. [24] proposed a profound transfer learning-based framework to discern COVID-19 cases and perform the segmentation of infections within chest X-ray images. This framework employs a two-stage cascaded architecture, incorporating a meticulously fine-tuned SegNet semantics, a segmentation-model classifier, and a SqueezeNet network. Utilizing chest X-ray images, Pan et al. [25] advocated a multi-channel featured deep neural network (MFDNN) method for the identification of individuals infected with COVID-19. To mitigate the inherent bias resulting from imbalanced data, the authors augmented the MFDNN model with data oversampling techniques during the training phase. Addressing the task of detecting COVID-19 cases through X-ray images encompassing normal, COVID-19, and viral pneumonia instances, Yang et al. [26] introduced CovidViT, an advanced transformer-based DL methodology termed COVID-Vision-Transformers. Suganyadevi and Seethalakshmi [27] introduced a classification methodology leveraging deep CNN models, namely COVID-HybridNetwork1 and COVID-HybridNetwork2. These models incorporate boundary-based techniques alongside convolutional procedures to classify X-ray images and regions affected by COVID-19. In the pursuit of COVID-19 and pneumonia identification from chest X-ray images, Bhattacharyya et al. [28] suggested a comprehensive pipeline. Their approach integrates segmented lung images, key-point extraction techniques, and conditional generative adversarial network.

The majority of existing works employ CNNs for COVID-19 detection [23]. In recent times, Convolutional Neural Networks (CNNs) have been surpassed in performance by a sophisticated architectural paradigm known as the transformer, which operates based on attention mechanisms [29], [30]. The fundamental underpinning of a transformer lies within a mechanism capable of assimilating the interdependencies existing between input and output sequences in a manner that obviates the necessity for recurrent iterations [31]. This makes the parallelization of transformer implementations simple and computationally efficient. Motivated by the triumph of transformers within the realm of NLP, Dosovitskiy et al. [29] devised the Vision Transformer (ViT), an adapted transformer framework specifically tailored to cater to the exigencies of computer vision. In order to undertake the task of image classification, Dosovitskiy et al. [29] made noteworthy modifications to the initial transformer model, enabling it to accommodate a sequence of image patches with fixed dimensions. These patches are handled in a manner akin to words commonly observed in NLP domains. Despite showing promise with natural images, there has been little research on ViTs in the research area of medical imaging, with the majority of studies using computed tomography (CT) [32]–[34] and chest radiography data [31].

This study posits a profound learning framework, namely the ViT model, as a means to accomplish the automated identification of COVID-19. The ViT model, fine-tuned for this purpose, boasts an all-encompassing architecture that eschews the utilization of any feature extraction methods, instead relying solely on unprocessed chest X-ray images to deliver diagnostic outcomes. To train this model, binary, ternary, and quaternary classifications are employed, leveraging diverse datasets substantiated in existing scholarly works. The following is a summary of the contributions of this work:

- A DL model was proposed utilizing X-ray imaging to facilitate the identification and diagnosis of individuals afflicted with COVID-19.
- The inquiry pertained to the transfer learning aptitude of transformers that underwent training on X-ray imagery.
- Employing discrete binary and ternary datasets alongside a quaternary dataset, the efficacy of the meticulously refined ViT models was assessed.
- Employing two discrete binary datasets, a remarkable accuracy rate of 99.92% and 99.84% was successfully attained.
- The accuracy for ternary classes was 97.95% and 86.48% for each dataset, and the accuracy for quaternary classes was 86.81%.
- The meticulously calibrated ViT model possesses the potential to assist medical practitioners in expediting and enhancing the accuracy of diagnostic procedures.

The structural arrangement of the article is outlined in the subsequent manner. Section 2 delivers a description of both binary and multi-class datasets together with an explanation of the ViT, and evaluation metrics. In section 3, experiments are conducted on various binary and multi-class datasets, and the outcomes are presented in terms of various evaluation metrics. In the final section, concluding remarks are stated.

## 2. Materials and Methods

### 2.1. Dataset Description

The utilization of datasets is indispensable in facilitating the execution of dependable ML experiments. Within this study, an assessment was conducted to gauge the efficacy of an ML model, employing X-ray datasets [1], [20]. These datasets play a crucial role in furnishing the requisite data for acquiring reliable outcomes. To thoroughly evaluate the performance of the scrutinized model, the datasets [1], [20], and [18], which were employed in earlier investigations, underwent binary, ternary, and quaternary classifications.

The initial dataset was formulated based on the extensive investigation conducted in the realm of scholarly inquiry outlined by reference [1]. This meticulously crafted dataset comprises three distinct subcategories of X-ray images, encompassing patients who have tested positive for COVID-19 or have been diagnosed with alternative viral and bacterial pneumonia conditions, juxtaposed with X-ray images depicting healthy control subjects. The COVID-19 instances were meticulously sourced from the work of Cohen et al. [35], an esteemed Kaggle competition [36], and the scholarly research documented within reference [1]. Notably, a comprehensive assemblage of 231 cases of COVID-19, confirmed via RT-PCR diagnosis, has been incorporated within the dataset. On the other hand, control samples representing normal physiological conditions, alongside cases of pneumonia attributable to bacterial pathogens (MERS, ARDS, SARS, and related etiologies), have been obtained from a solitary publicly accessible resource accessible through the Mendeley data website [37], [38]. Within this repository, a notable sum of 2,780 cases characterized by bacterial pneumonia, alongside 1,493 cases pertaining to viral pneumonia, and a further 1,583 instances constituting normal control cases, have been successfully accumulated.

The second dataset was formulated as the foundational groundwork for the investigation outlined in reference [18]. Within this dataset, X-ray images obtained from two distinct sources as mentioned in reference [18] were employed for the purpose of COVID-19 identification. Cohen JP [39] undertook the development of a dedicated COVID-19 X-ray image dataset by aggregating images from multiple publicly accessible sources. Notably, a total of 125 X-ray images were duly diagnosed with COVID-19 throughout the study duration documented in reference [18]. Furthermore, the Chest X-ray dataset devised by Wang et al. [40] was judiciously employed to incorporate images representing pneumonia cases and those depicting normal conditions. In order to address the issue of data imbalance, a systematic random selection process was employed to procure a total of 500 pneumonia-afflicted images and an equal number of healthy-grade anterior chest radiography images from this aforementioned dataset.

### 2.2. Vision Transformer

Image transformer architectures, exemplified by the ViT as proposed in scholarly works [29] and [41], deploy a multi-headed self-attention mechanism with the aim of transforming non-sequential spatial signals, particularly images, into a sequence of diminutive, unchanging patches of fixed dimensions (e.g. 32x32). This mechanism facilitates the extraction of comprehensive contextual information from the input image. Moreover, by employing multiple heads in the self-attention mechanism, the model becomes capable of selectively attending to distinct regions within the image and comprehending the interdependencies over extended temporal spans among various patches. By adopting this approach, the model effectively mitigates the inherent biases originating from tensor-based operations, such as locally constrained receptive fields and translation invariance, which are commonly encountered in conventional CNNs. The schematic representation of the ViT model can be observed in Figures 1 and 2.

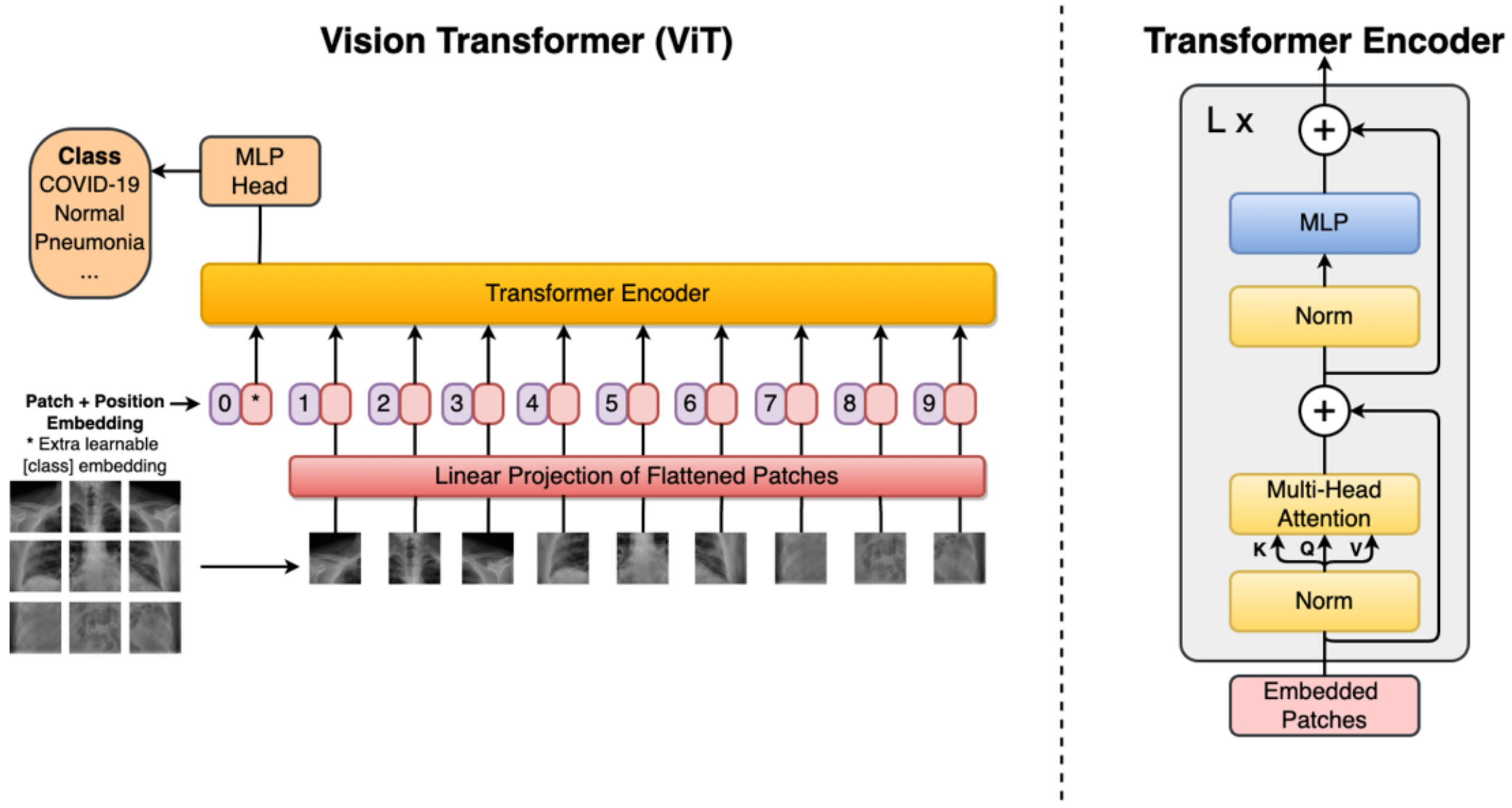

Fig. 1. The overall architecture of ViT [29].

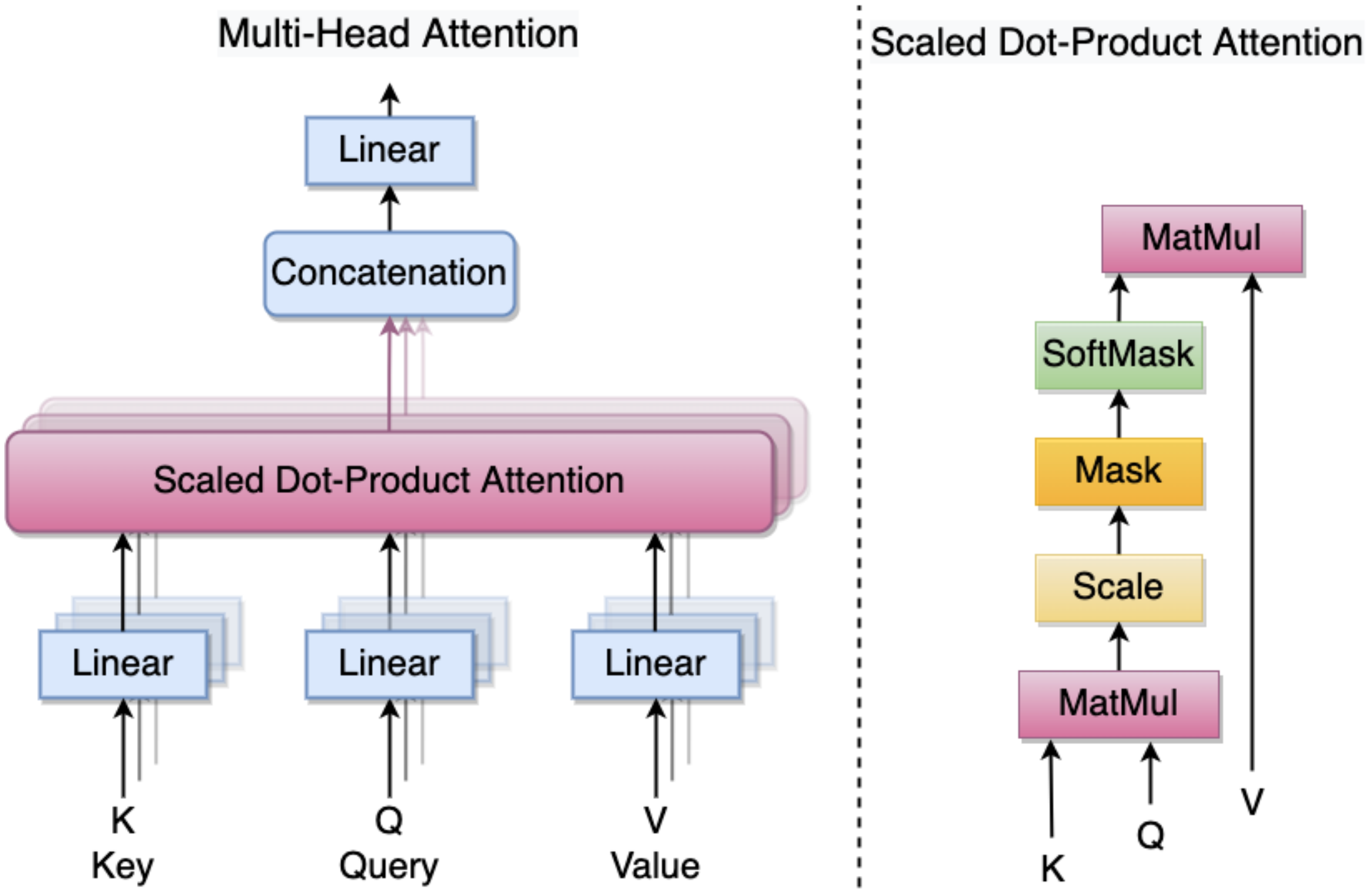

Fig. 2. The internal structure of Multi-Head Attention [29][41]

ViT begins by reshaping the input ($x \in R^{H \times W \times C}$; $C$ = number of channels, and $H \times W$ = image resolution) into a series of 2-dimensional patches ($\{x_p^i \in R^{p^2 \cdot C} | i = 1| \ldots |N\}$; $N = \frac{H \times W}{P^2}$ and $patch\ size = P \times P$). Subsequently, employing trainable linear projections, the vectorized patches undergo a linear mapping process, wherein they are transformed into a latent embedding space of $D$ dimensions ($E \in R^{(p^2 \cdot C) \times D}$; D = constant latent vector size). Alongside the incorporation of patch embeddings, the inclusion of positional embeddings ($E_{pos} \in R^{(N+1) \times D}$) serves to encode the spatial information inherent in the input images – see Eq. 1:

$$Z_0 = [x_p^1 E; x_p^2 E; \ldots; x_p^N E] + E_{pos} \tag{1}$$

In the context of the transformer encoder, the utilization of multilayer perceptron (MLP; refer to Eq. 3) blocks and a specified quantity ($H$) of self-attention heads (MSA; refer to Eq. 2) serves the purpose of generating the encoded image representation. Finally, aggregated information and the inter-patch dependencies are supplied to an MLP head, which makes a decision based on the input [42]:

$$z_l = MLP(LN(z_l')) + z_l' \tag{2}$$

$$z_l' = MSA(LN(z_{l-1})) + z_{l-1} \tag{3}$$

where $z_l \in \mathbf{R}^{\frac{HW}{p^2} \times D}$ symbolizes the encoded feature representation, while LN(.) refers to the operator for layer normalization. The equation that encapsulates a single self-attention head can be expressed as Eq. 4:

$$\text{Attention}_h(\text{X}) = \sigma\left(\frac{\text{QK}^\text{T}}{\sqrt{d_k}}\right)\text{V} \tag{4}$$

where σ is the Softmax function; V (value) = $XW_v$; K (key) = $XW_k$; Q (query) = $XW_q$; $W_v$, $W_k$, and $W_q$, represent learnable matrices; $d_k$ is the dimension of K, and T denotes matrix transpose.

## 2.3. Evaluation metrics

In order to measure the generalizability of a proposed model, its performance on test data serves as a pivotal indicator. In the realm of binary classification, performance is evaluated through the utilization of four metrics. Firstly, accuracy is employed to gauge the proportion of correctly classified test instances, as denoted by Eq. 5. This metric has been widely employed in the realm of COVID-19 diagnosis, as evidenced by various studies [5], [20], [43]. Secondly, sensitivity or recall is utilized to assess the proportion of COVID-19 cases accurately detected, as expressed in Eq. 7. Thirdly, specificity is employed to determine the proportion of non-COVID-19 cases that were accurately excluded from classification. Lastly, the F1 score, defined as the harmonic mean of precision and recall components (as shown in Eq. 9), is leveraged as a comprehensive evaluation metric. Furthermore, the area under the receiver operating characteristic curve (ROC), known as AUC, represents a holistic assessment that carefully considers both sensitivity and specificity. In multi-class classification, accuracy serves as the primary metric, as demonstrated by several recent studies [5], [20], [31]. Additionally, macro-averaged precision, recall, and F1-score are commonly employed. Here, FN denotes False Negatives, FP denotes False Positives, TN denotes True Negatives, and TP denotes True Positives.

$$Accuracy = \frac{TP + TN}{TP + TN + FP + FN} \tag{5}$$

$$Precision = \frac{TP}{TP + FP} \tag{6}$$

$$Recall\ (Sensitivity) = \frac{TP}{TP + FN} \tag{7}$$

$$Specificity = \frac{TN}{TN + FP} \tag{8}$$

$$F1\ Score = 2 \times \frac{Precision \times Recall}{Precision + Recall} \tag{9}$$

# 3. Experiments

### 3.1. Experimental setup

In the present study, an architectural framework known as the ViT model was employed, more specifically the ViT-base model comprising 12 layers. The model encompasses a hidden size of 768, an MLP size of 3072, 12 attention heads, and an impressive parameter count of 86 million. To adapt the model to our specific task, the process of fine-tuning was undertaken, capitalizing on the pre-training accomplished with the extensive ImageNet-21k dataset. This paradigm of transfer learning was employed as a means to enhance the development of our model. The process entailed initializing the model with the pre-trained weights, substituting the classification head with a novel head tailored to our dataset's distinct classes, which could be binary, ternary, or quaternary in nature. Furthermore, the network parameters were frozen, thereby preserving the knowledge acquired during pre-training, and allowing focused training to take place. Subsequently, the model underwent refinement by integrating image augmentation techniques, facilitating the augmentation of the dataset and ultimately enhancing its overall performance.

Image augmentation is a widely employed technique for the creation of robust image classifiers when faced with a scarcity of training data. Within the context of this study, the data underwent image augmentation procedures involving random cropping to achieve image dimensions of 224x224 pixels. Furthermore, horizontal flipping of the resized image was performed. Moreover, normalization of the images was executed utilizing the ImageNet21k values, characterized by a standard deviation of [0.229, 0.224, 0.225] and a mean of [0.485, 0.456, 0.406].

The AdamW optimizer [44] was employed to optimize all networks, utilizing a learning rate of 0.00002 and a batch size of 8. These parameter values were meticulously determined through a process of rigorous experimentation. The optimal model, exhibiting superior accuracy, was meticulously selected based on its performance following the completion of 50 epochs of network training. The training process was executed on the Google Colab Pro platform, which provided ample computational resources in the form of 30 GB RAM, complemented by the utilization of the Nvidia Tesla P100 with 16 GB VRAM.

To appraise the efficacy of the suggested framework in tackling binary as well as multi-classification quandaries, a rigorous 5-fold cross-validation methodology was employed. A distinct subset of X-ray images, constituting 20% of the overall dataset, was exclusively reserved for the purpose of validation, while the remaining 80% was allocated for training the model. This entire procedure was reiterated five times, ensuring resilience across every fold. During the validation phase, the individual k partitions were meticulously folded and intricately integrated within the folds. The model's performance on the validation sets served as a pivotal yardstick for evaluating its efficacy.

### 3.2. Results and Discussion

*Experiments on Binary Classification Tasks*

The process of diagnosing COVID-19 through the utilization of X-ray images encompasses a dichotomous classification task, wherein the images are classified as either positive or negative for COVID-19. In order to estimate the effectiveness of the fine-tuned ViT model, an experimental assessment was conducted on datasets characterized by a binary nature. The foundational framework for this assessment involved employing the pre-trained ImageNet-21k model. The integration of a classifier head onto the ViT architecture facilitated the utilization of the ViT model for the purpose of discerning chest X-ray images, effectively distinguishing them into distinct classes of COVID-19 and non-COVID-19. In order to gauge the efficacy of the model, a meticulous and comprehensive evaluation spanning 50 training epochs was conducted, encompassing the computation of crucial metrics such as sensitivity, specificity, precision, F1-score, accuracy, and the AUC. The resulting outcomes derived from this comprehensive evaluation are meticulously outlined and presented in Table 1.

Table 1. The evaluation outcomes pertaining to the proposed model utilized for binary classification

| **Binary Dataset** | **Folds** | **Performance Metrics** | | | | | |
|---|---|---|---|---|---|---|---|
| | | **Accuracy** | **Precision** | **Sensitivity** | **F1-score** | **Specificity** | **AUC** |
| The dataset used by Ozturk et al. [18] | 1 | 1 | 1 | 1 | 1 | 1 | 1 |
| | 2 | 1 | 1 | 1 | 1 | 1 | 1 |
| | 3 | 1 | 1 | 1 | 1 | 1 | 1 |
| | 4 | 0.984 | 1 | 0.92 | 0.9583 | 1 | 0.9992 |
| | 5 | 0.976 | 1 | 0.88 | 0.9361 | 1 | 0.9984 |
| | **Average** | 0.992 | 1 | 0.96 | 0.9789 | 1 | 0.9995 |
| | | | | | | | |
| The dataset used by Li et al. [1] | 1 | 0.9984 | 0.9915 | 0.9915 | 0.9915 | 0.9991 | 0.9998 |
| | 2 | 1 | 1 | 1 | 1 | 1 | 1 |
| | 3 | 0.9961 | 0.975 | 0.9831 | 0.979 | 0.9974 | 0.9979 |
| | 4 | 0.9984 | 1 | 0.9831 | 0.9915 | 1 | 0.9984 |
| | 5 | 0.9992 | 1 | 0.9915 | 0.9957 | 1 | 0.9999 |
| | **Average** | 0.9984 | 0.9933 | 0.9898 | 0.9915 | 0.9993 | 0.9992 |

*Experiments on Multi-class Classification Tasks*

This study entails the utilization of two distinct datasets to undertake a multi-class classification task, encompassing ternary and quaternary classes. The experimental assessment of the proposed ViT model for multi-class datasets was executed employing the pre-trained ImageNet-21k model as the foundational framework. The ViT model was deployed to discern classifications of chest X-ray images into three or four classes, employing a transfer learning strategy that entailed incorporating a classifier head atop the ViT. Following training, the model's performance was assessed by computing accuracy, precision, sensitivity, and F1-score. The ensuing evaluation outcomes are meticulously outlined in Table 2.

Table 2. The evaluation outcomes pertaining to the proposed model utilized for multi-class classification

| **Multi-class Dataset** | **Folds** | **Performance Metrics** | | | |
|---|---|---|---|---|---|
| | | **Accuracy** | **Precision** | **Sensitivity** | **F1-score** |
| The dataset used by Ozturk et al. [18] (Three classes) | 1 | 0.8444 | 0.8867 | 0.8233 | 0.8494 |
| | 2 | 0.88 | 0.9146 | 0.83 | 0.8616 |
| | 3 | 0.8311 | 0.8815 | 0.8533 | 0.8619 |
| | 4 | 0.8977 | 0.9258 | 0.8733 | 0.8953 |
| | 5 | 0.8711 | 0.8983 | 0.8633 | 0.8770 |
| | **Average** | 0.8648 | 0.9014 | 0.8486 | 0.8690 |
| | | | | | |
| The dataset used by Li et al. [1] (Three classes) | 1 | 0.9767 | 0.9652 | 0.9794 | 0.9721 |
| | 2 | 0.9767 | 0.9702 | 0.9766 | 0.9733 |
| | 3 | 0.9813 | 0.9762 | 0.9853 | 0.9807 |
| | 4 | 0.9852 | 0.9848 | 0.9861 | 0.9854 |
| | 5 | 0.9775 | 0.972 | 0.9789 | 0.9758 |
| | **Average** | 0.9795 | 0.9739 | 0.9813 | 0.9775 |
| | | | | | |
| Custom Dataset (Four classes) | 1 | 0.8596 | 0.8752 | 0.8692 | 0.8696 |
| | 2 | 0.8674 | 0.8776 | 0.8855 | 0.8813 |
| | 3 | 0.8643 | 0.8818 | 0.8833 | 0.8824 |
| | 4 | 0.8713 | 0.8871 | 0.8810 | 0.8830 |
| | 5 | 0.8782 | 0.8957 | 0.8926 | 0.8935 |
| | **Average** | 0.8681 | 0.8835 | 0.8823 | 0.8820 |

The proposed model underwent a comprehensive evaluation encompassing various state-of-the-art techniques employed for both binary and multi-class classification endeavors. The comparative analysis in Table 3 showcases the attainment of binary classification results by the proposed model, juxtaposed against other cutting-edge techniques for the classification of X-ray images into COVID-19 and non-COVID-19 classes. Subsequently, Table 4 concisely encapsulates the outcomes derived from the proposed model and alternative state-of-the-art methodologies employed for the multi-class classification task involving diverse image classes discerned from X-ray images. It is of paramount significance to acknowledge that this study deliberately excluded modalities such as CT scans and ultrasound due to their distinct characteristics and potential ramifications on the classification outcomes. The results presented in Tables 3 and 4 incontrovertibly substantiate the efficacy of the proposed model in both binary and multi-class classification undertakings.

Table 3. A comparative analysis of the binary classification outcomes attained by the proposed model in contrast to those obtained by other state-of-the-art methodologies (%).

| **Reference** | **Approach** | **Image count** | **Accuracy** | **Precision** | **Sensitivity** | **F1-Score** | **Specificity** | **AUC** |
|---|---|---|---|---|---|---|---|---|
| ***This study*** | ViT | 125 COVID-19 vs. 500 No Findings | 99.2 | 100 | 96 | 97.89 | 100 | 97.48 |
| | ViT | 231 COVID-19 vs. 1583 Normal | 99.84 | 99.33 | 98.98 | 99.15 | 99.93 | 99.92 |
| Li et al. [1] | Multi-task contrastive learning | 231 COVID-19 vs. 1583 Normal | 97.23 | - | 92.97 | - | 91.91 | 92.13 |
| Ozturk et al. [18] | CNN | 125 COVID-19 vs. 500 No Findings | 98.08 | 98.03 | 95.13 | 96.51 | 95.3 | |
| Hassantabar et al. [17] | CNN | 315 COVID-19 vs. 367 non-COVID-19 | 93.20 | - | 96.10 | - | - | - |
| Pandit et al. [19] | VGG-16 | 224 COVID-19 vs. 1204 non-COVID-19 | 96.0 | - | 92.64 | - | 97.27 | - |
| Ismael and Sengur et al. [21] | Resnet50 with SVM | 180 COVID-19 vs. 200 non-COVID-19 | 94.70 | - | - | - | - | - |
| Gayathri et al. [23] | CNN with neural network for classification and AE for dimensionality reduction | 504 COVID-19 vs. 542 non-COVID-19 | 95.78 | 0.9563 | 95.63 | 95.63 | 95.94 | 98.21 |
| Sundaram et al. [24] | Residual SqueezeNet and SegNet | 700 COVID-19 1400 Non-COVID-19 | 99.38 | - | - | - | - | - |
| Rashid et al. [22] | AutoCovNet | 408 COVID-19 and 816 Non-COVID-19 | 99.39 | - | 99.39 | - | 100 | - |

The assessment of the proposed model was conducted through a comprehensive analysis of multiple studies employing diverse approaches. Each of these studies employed a distinct number of instances for training the model. The evaluation process employed in this research entailed the utilization of a held-out validation set, as opposed to the cross-validation approach utilized by other methodologies, as observed in works such as Pandit et al. [19] and Ismael and Sengur et al. [21].

Despite the substantial outcomes achieved by all the methodologies scrutinized in this investigation, the proposed model demonstrated a commensurate level of accuracy in the multi-class classification task when compared to the approach employed in [1]. Furthermore, the outcomes of the binary classification, in terms of sensitivity, precision, F1-score, and accuracy, were found to be comparable to the approach employed in [18]. It is worth noting, however, that the utilization of varying numbers of images in each methodology prevents a fair comparison of the results.

Table 4. A comparative analysis of the multi-class classification outcomes attained by the proposed model in contrast to those obtained by other state-of-the-art methodologies (%).

| **Reference** | **Approach** | **Image count** | **Accuracy** | **Precision** | **Sensitivity** | **F1-score** |
|---|---|---|---|---|---|---|
| ***This study*** | ViT | 231 COVID-19, 4007 pneumonia, 1583 normal [1] | 97.95 | 97.39 | 98.13 | 97.75 |
| | ViT | 500 No findings, 500 pneumonia, and 125 Covid-19 [18] | 86.48 | 90.14 | 84.86 | 86.90 |
| | ViT | 1583 Normal, 2780 bacterial pneumonia, 1493 viral pneumonia, 593 COVID-19 | 86.81 | 88.35 | 88.23 | 88.20 |
| Li et al. [1] | Multi-task contrastive learning | 231 COVID-19, 4007 pneumonia, 1583 normal | 93.49 | - | - | - |
| Apostolopoulos et al. [20] | MobileNet, Xception, Inception ResNetv2, Inception, and VGG-19 | 224 Covid-19, 504 healthy, and 714 pneumonia | 94.72 | - | - | - |
| Sundaram et al. [24] | Residual SqueezeNet and SegNet | 221 COVID-19, 700 Viral Pneumonia 700 Normal | 99.14 | - | - | - |
| Yang et al. [26] | CovidViT | 3616 COVID-19, 10192 Normal, 1345 Viral Peumonia | 98.2 | 97.8 | 95.7 | 95.3 |
| Rashid et al. [22] | AutoCovNet | 408 COVID-19, 408 Normal, 408 Pneumonia | 96.45 | - | 95.94 | |
| Ozturk et al. [18] | CNN | 500 No findings, 500 pneumonia, and 125 Covid-19 | 87.02 | 89.96 | 85.35 | 87.37 |

## 4. Conclusion

In summary, this investigation offers a comprehensive assessment of the performance of a ViT model in the context of binary and multi-class task classification utilizing X-ray imagery. The proposed model demonstrates outstanding accuracy and encompasses a wide range of performance metrics across both multi-class and binary classification scenarios.

Regarding the binary classification task, the proposed ViT model displays a remarkable ability to accurately discern non-COVID-19 and COVID-19 images, achieving accuracies of 99.2% and 99.84% respectively. In the realm of multi-class classification, the proposed ViT model demonstrates exceptional efficacy in the allocation of images to three or four discrete classes, resulting in remarkable accuracies of 97.95% and 86.81% correspondingly.

Compared to other cutting-edge algorithms, which also exhibit noteworthy accuracy and performance metrics, a meticulous comparison reveals the superior efficacy of the proposed ViT model in terms of accuracy, sensitivity, and F1-score within the domain of multi-class classification. The outcomes of this empirical investigation conclusively establish the feasibility of employing ViT models for the purpose of discriminating COVID-19 and non-COVID-19 X-ray images.

Expanding the analysis to encompass alternative imaging modalities, namely CT scans and ultrasound, would yield valuable insights into the efficacy of the proposed ViT model across diverse modalities, thereby amplifying the breadth of this study. Subsequent research endeavors could attempt to enhance the performance of the suggested model by training it from scratch. Additionally, an intriguing avenue of investigation would involve evaluating the performance of the suggested model within an authentic clinical environment, as it would facilitate an assessment of its diagnostic potential for COVID-19.

**Data Availability**

The dataset employed in this study was acquired from antecedently published scholarly works.

**Competing interests**

None Declared

**Funding**

Not applicable

**Declaration of generative AI and AI-assisted technologies in the writing process**

During the preparation of this work the author used ChatGPT in order to polish the article text. After using this tool/service, the author reviewed and edited the content as needed and takes full responsibility for the content of the publication.

## References

[1] J. Li *et al.*, “Multi-task contrastive learning for automatic CT and X-ray diagnosis of COVID-19,” *Pattern Recognit*, vol. 114, p. 107848, Jun. 2021, doi: 10.1016/J.PATCOG.2021.107848

[2] W. Guan *et al.*, "Clinical Characteristics of Coronavirus Disease 2019 in China," *New England Journal of Medicine*, vol. 382, no. 18, pp. 1708–1720, Apr. 2020, doi: 10.1056/NEJMOA2002032/SUPPL_FILE/NEJMOA2002032_DISCLOSURES.PDF. Available: https://www.nejm.org/doi/10.1056/NEJMoa2002032. [Accessed: Aug. 18, 2022]

[3] C. Huang *et al.*, "Clinical features of patients infected with 2019 novel coronavirus in Wuhan, China," *The Lancet*, vol. 395, no. 10223, pp. 497–506, Feb. 2020, doi: 10.1016/S0140-6736(20)30183-5

[4] Z. Xu *et al.*, "Pathological findings of COVID-19 associated with acute respiratory distress syndrome," *Lancet Respir Med*, vol. 8, no. 4, pp. 420–422, Apr. 2020, doi: 10.1016/S2213-2600(20)30076-X

[5] X. Xie *et al.*, "An Infectious cDNA Clone of SARS-CoV-2," *Cell Host Microbe*, vol. 27, no. 5, pp. 841-848.e3, May 2020, doi: 10.1016/J.CHOM.2020.04.004

[6] G. Litjens *et al.*, "A survey on deep learning in medical image analysis," *Med Image Anal*, vol. 42, pp. 60–88, Dec. 2017, doi: 10.1016/J.MEDIA.2017.07.005

[7] O. Faust, Y. Hagiwara, T. J. Hong, O. S. Lih, and U. R. Acharya, "Deep learning for healthcare applications based on physiological signals: A review," *Comput Methods Programs Biomed*, vol. 161, pp. 1–13, Jul. 2018, doi: 10.1016/J.CMPB.2018.04.005

[8] F. Murat, O. Yildirim, M. Talo, U. B. Baloglu, Y. Demir, and U. R. Acharya, "Application of deep learning techniques for heartbeats detection using ECG signals-analysis and review," *Comput Biol Med*, vol. 120, p. 103726, May 2020, doi: 10.1016/J.COMPBIOMED.2020.103726

[9] Y. Lecun, Y. Bengio, and G. Hinton, "Deep learning," *Nature 2015 521:7553*, vol. 521, no. 7553, pp. 436–444, May 2015, doi: 10.1038/nature14539. Available: https://www.nature.com/articles/nature14539. [Accessed: Aug. 18, 2022]

[10] "ImageNet Classification with Deep Convolutional Neural Networks." Available: https://proceedings.neurips.cc/paper/2012/hash/c399862d3b9d6b76c8436e924a68c45b-Abstract.html. [Accessed: Aug. 18, 2022]

[11] S. Parvaneh, J. Rubin, S. Babaeizadeh, and M. Xu-Wilson, "Cardiac arrhythmia detection using deep learning: A review," *J Electrocardiol*, vol. 57, pp. S70–S74, Nov. 2019, doi: 10.1016/J.JELECTROCARD.2019.08.004

[12] A. Kursad Poyraz, S. Dogan, E. Akbal, and T. Tuncer, "Automated brain disease classification using exemplar deep features," *Biomed Signal Process Control*, vol. 73, p. 103448, Mar. 2022, doi: 10.1016/J.BSPC.2021.103448

[13] S. S. Koshy, L. J. Anbarasi, M. Jawahar, and V. Ravi, "Breast cancer image analysis using deep learning techniques – a survey," *Health Technol (Berl)*, vol. 12, no. 6, pp. 1133–1155, Nov. 2022, doi: 10.1007/S12553-022-00703-5/FIGURES/8. Available: https://link.springer.com/article/10.1007/s12553-022-00703-5. [Accessed: Jan. 17, 2023]

[14] P. Malhotra, S. Gupta, D. Koundal, A. Zaguia, M. Kaur, and H. N. Lee, "Deep Learning-Based Computer-Aided Pneumothorax Detection Using Chest X-ray Images," *Sensors 2022, Vol. 22, Page 2278*, vol. 22, no. 6, p. 2278, Mar. 2022, doi: 10.3390/S22062278. Available: https://www.mdpi.com/1424-8220/22/6/2278/htm. [Accessed: Aug. 18, 2022]

[15] C. di Ruberto *et al.*, "Pneumonia Detection on Chest X-ray Images Using Ensemble of Deep Convolutional Neural Networks," *Applied Sciences 2022, Vol. 12, Page 6448*, vol. 12, no. 13, p. 6448, Jun. 2022, doi: 10.3390/APP12136448. Available: https://www.mdpi.com/2076-3417/12/13/6448/htm. [Accessed: Aug. 17, 2022]

[16] F. Caobelli, “Artificial intelligence in medical imaging: Game over for radiologists?,” *Eur J Radiol*, vol. 126, May 2020, doi: 10.1016/J.EJRAD.2020.108940. Available: http://www.ejradiology.com/article/S0720048X20301297/fulltext. [Accessed: Aug. 18, 2022]

[17] S. Hassantabar, M. Ahmadi, and A. Sharifi, “Diagnosis and detection of infected tissue of COVID-19 patients based on lung x-ray image using convolutional neural network approaches,” *Chaos Solitons Fractals*, vol. 140, p. 110170, Nov. 2020, doi: 10.1016/J.CHAOS.2020.110170

[18] T. Ozturk, M. Talo, E. A. Yildirim, U. B. Baloglu, O. Yildirim, and U. Rajendra Acharya, “Automated detection of COVID-19 cases using deep neural networks with X-ray images,” *Comput Biol Med*, vol. 121, p. 103792, Jun. 2020, doi: 10.1016/J.COMPBIOMED.2020.103792

[19] M. K. Pandit, S. A. Banday, R. Naaz, and M. A. Chishti, “Automatic detection of COVID-19 from chest radiographs using deep learning,” *Radiography*, vol. 27, no. 2, pp. 483–489, May 2021, doi: 10.1016/J.RADI.2020.10.018

[20] I. D. Apostolopoulos and T. A. Mpesiana, “Covid-19: automatic detection from X-ray images utilizing transfer learning with convolutional neural networks,” *Phys Eng Sci Med*, vol. 43, no. 2, pp. 635–640, Jun. 2020, doi: 10.1007/S13246-020-00865-4/TABLES/6. Available: https://link.springer.com/article/10.1007/s13246-020-00865-4. [Accessed: Aug. 18, 2022]

[21] A. M. Ismael and A. Şengür, “Deep learning approaches for COVID-19 detection based on chest X-ray images,” *Expert Syst Appl*, vol. 164, p. 114054, Feb. 2021, doi: 10.1016/J.ESWA.2020.114054

[22] N. Rashid, M. A. F. Hossain, M. Ali, M. Islam Sukanya, T. Mahmud, and S. A. Fattah, “AutoCovNet: Unsupervised feature learning using autoencoder and feature merging for detection of COVID-19 from chest X-ray images,” *Biocybern Biomed Eng*, vol. 41, no. 4, pp. 1685–1701, Oct. 2021, doi: 10.1016/J.BBE.2021.09.004

[23] J. L. Gayathri, B. Abraham, M. S. Sujarani, and M. S. Nair, “A computer-aided diagnosis system for the classification of COVID-19 and non-COVID-19 pneumonia on chest X-ray images by integrating CNN with sparse autoencoder and feed forward neural network,” *Comput Biol Med*, vol. 141, p. 105134, Feb. 2022, doi: 10.1016/J.COMPBIOMED.2021.105134

[24] S. G. Sundaram, S. A. Aloyuni, R. A. Alharbi, T. Alqahtani, M. Y. Sikkandar, and C. Subbiah, “Deep Transfer Learning Based Unified Framework for COVID19 Classification and Infection Detection from Chest X-Ray Images,” *Arab J Sci Eng*, vol. 47, no. 2, pp. 1675–1692, Feb. 2022, doi: 10.1007/S13369-021-05958-0/TABLES/8. Available: https://link.springer.com/article/10.1007/s13369-021-05958-0. [Accessed: Jan. 17, 2023]

[25] L. Pan *et al.*, “MFDNN: multi-channel feature deep neural network algorithm to identify COVID19 chest X-ray images,” *Health Inf Sci Syst*, vol. 10, no. 1, pp. 1–10, Dec. 2022, doi: 10.1007/S13755-022-00174-Y/FIGURES/6. Available: https://link.springer.com/article/10.1007/s13755-022-00174-y. [Accessed: Jan. 17, 2023]

[26] H. Yang, L. Wang, Y. Xu, and X. Liu, “CovidViT: a novel neural network with self-attention mechanism to detect Covid-19 through X-ray images,” *International Journal of Machine Learning and Cybernetics*, pp. 1–15, Oct. 2022, doi: 10.1007/S13042-022-01676-7/TABLES/11. Available: https://link.springer.com/article/10.1007/s13042-022-01676-7. [Accessed: Jan. 17, 2023]

[27] S. Suganyadevi and V. Seethalakshmi, “CVD-HNet: Classifying Pneumonia and COVID-19 in Chest X-ray Images Using Deep Network,” *Wirel Pers Commun*, vol. 126, no. 4, pp. 3279–3303, Oct. 2022, doi: 10.1007/S11277-022-09864-Y/TABLES/5. Available: https://link.springer.com/article/10.1007/s11277-022-09864-y. [Accessed: Jan. 17, 2023]

[28] A. Bhattacharyya, D. Bhaik, S. Kumar, P. Thakur, R. Sharma, and R. B. Pachori, “A deep learning based approach for automatic detection of COVID-19 cases using chest X-ray images,” *Biomed Signal Process Control*, vol. 71, p. 103182, Jan. 2022, doi: 10.1016/J.BSPC.2021.103182

[29] A. Dosovitskiy *et al.*, “An Image is Worth 16x16 Words: Transformers for Image Recognition at Scale,” Oct. 2020, Available: https://arxiv.org/abs/2010.11929v2. [Accessed: Mar. 02, 2022]

[30] A. Vaswani *et al.*, “Attention Is All You Need”, doi: 10.5555/3295222.3295349

[31] M. Usman, T. Zia, and A. Tariq, “Analyzing Transfer Learning of Vision Transformers for Interpreting Chest Radiography,” *Journal of Digital Imaging 2022*, pp. 1–18, Jul. 2022, doi: 10.1007/S10278-022-00666-Z. Available: https://link.springer.com/article/10.1007/s10278-022-00666-z. [Accessed: Aug. 24, 2022]

[32] Y. Dai, Y. Gao, and F. Liu, “TransMed: Transformers Advance Multi-modal Medical Image Classification,” *Diagnostics*, vol. 11, no. 8, Mar. 2021, doi: 10.48550/arxiv.2103.05940. Available: https://arxiv.org/abs/2103.05940v1. [Accessed: Aug. 24, 2022]

[33] X. Gao, Y. Qian, and A. Gao, “COVID-VIT: Classification of COVID-19 from CT chest images based on vision transformer models,” Jul. 2021, doi: 10.48550/arxiv.2107.01682. Available: https://arxiv.org/abs/2107.01682v1. [Accessed: Aug. 24, 2022]

[34] N. Ghassemi *et al.*, “Automatic Diagnosis of COVID-19 from CT Images using CycleGAN and Transfer Learning,” Apr. 2021, doi: 10.48550/arxiv.2104.11949. Available: https://arxiv.org/abs/2104.11949v1. [Accessed: Aug. 24, 2022]

[35] “ieee8023/covid-chestxray-dataset: We are building an open database of COVID-19 cases with chest X-ray or CT images.” Available: https://github.com/ieee8023/covid-chestxray-dataset. [Accessed: Aug. 23, 2022]

[36] “COVID-19 X rays | Kaggle.” Available: https://www.kaggle.com/datasets/andrewmvd/convid19-x-rays. [Accessed: Aug. 23, 2022]

[37] D. Kermany, K. Zhang, and M. Goldbaum, “Large Dataset of Labeled Optical Coherence Tomography (OCT) and Chest X-Ray Images,” vol. 3, 2018, doi: 10.17632/RSCBJBR9SJ.3

[38] D. S. Kermany *et al.*, “Identifying Medical Diagnoses and Treatable Diseases by Image-Based Deep Learning,” *Cell*, vol. 172, no. 5, pp. 1122-1131.e9, Feb. 2018, doi: 10.1016/J.CELL.2018.02.010/ATTACHMENT/CCBE1548-5469-4F17-BFDD-93EADB96C5E9/MMC1.PDF. Available: http://www.cell.com/article/S0092867418301545/fulltext. [Accessed: Jul. 03, 2022]

[39] J. P. Cohen, P. Morrison, and L. Dao, “COVID-19 Image Data Collection,” Mar. 2020, doi: 10.48550/arxiv.2003.11597. Available: https://arxiv.org/abs/2003.11597v1. [Accessed: Aug. 18, 2022]

[40] X. Wang, Y. Peng, L. Lu, Z. Lu, M. Bagheri, and R. M. Summers, “ChestX-ray8: Hospital-scale Chest X-ray Database and Benchmarks on Weakly-Supervised Classification and Localization of Common Thorax Diseases”, Available: https://uts.nlm.nih.gov/metathesaurus.html. [Accessed: Aug. 24, 2022]

[41] E. Asadi Shamsabadi, C. Xu, A. S. Rao, T. Nguyen, T. Ngo, and D. Dias-da-Costa, “Vision transformer-based autonomous crack detection on asphalt and concrete surfaces,” *Autom Constr*, vol. 140, p. 104316, Aug. 2022, doi: 10.1016/J.AUTCON.2022.104316

[42] K. Lee *et al.*, “ViTGAN: Training GANs with Vision Transformers,” Jul. 2021, doi: 10.48550/arxiv.2107.04589. Available: https://arxiv.org/abs/2107.04589v1. [Accessed: Aug. 24, 2022]

[43] L. Li *et al.*, “Using Artificial Intelligence to Detect COVID-19 and Community-acquired Pneumonia Based on Pulmonary CT: Evaluation of the Diagnostic Accuracy,” *Radiology*, vol. 296, no. 2, pp. E65–E71, Aug. 2020, doi: 10.1148/RADIOL.2020200905. Available: https://pubmed.ncbi.nlm.nih.gov/32191588/. [Accessed: Aug. 23, 2022]

[44] I. Loshchilov and F. Hutter, “Decoupled Weight Decay Regularization,” *7th International Conference on Learning Representations, ICLR 2019*, Nov. 2017, doi: 10.48550/arxiv.1711.05101. Available: https://arxiv.org/abs/1711.05101v3. [Accessed: Aug. 25, 2022]